\documentclass[pra,aps,superscriptaddress,floatfix,tightenlines,twocolumn,showpacs]{revtex4-1}
\usepackage{amsmath}
\usepackage{amssymb}
\usepackage{amsfonts}
\usepackage{dcolumn}
\usepackage{epsfig}
\usepackage{subfigure}
\usepackage[dvips]{color}
\usepackage{bm}
\usepackage{times}
\usepackage{amsthm}
\newcommand{\tr}[1]{\mbox{$\textrm{Tr}{\left(#1\right)}$}}

\begin{document}
\title{Quantum Metrological Bounds for Vector Parameter}
\author{Yu-Ran Zhang}
\affiliation{Beijing National Laboratory for Condensed Matter Physics, Institute
of Physics, Chinese Academy of Sciences, Beijing 100190, China}
\author{Heng Fan}
\email{hfan@iphy.ac.cn}
\affiliation{Beijing National Laboratory for Condensed Matter Physics, Institute of Physics, Chinese Academy of Sciences, Beijing 100190, China}
\affiliation{Collaborative Innovation Center of Quantum Matter, Beijing, China}
\date{\today}
\begin{abstract}
Precise measurement is crucial to science and technology.
However, the rule of nature imposes various restrictions
on the precision that can be achieved depending on specific methods
of measurement. In particular, quantum mechanics poses
the ultimate limit on precision which can only be approached but never
be violated. Depending on analytic techniques,
these bounds may not be unique.
Here, in view of prior information, we investigate systematically the precision
bounds of the total mean-square error of vector parameter estimation which contains $d$ independent
parameters. From quantum Ziv-Zakai error bounds, we derive two kinds of quantum metrological
bounds for vector parameter estimation, both of which should be satisfied. By these bounds, we
show that a constant advantage can be expected via simultaneous estimation strategy over
the optimal individual estimation strategy, which solves a long-standing problem.
A general framework for obtaining the lower bounds in 
a noisy system is also proposed.
\end{abstract}
\maketitle

\section{Introduction}
Quantum parameter estimation, the emerging field of quantum technology, aims to use
entanglement and other quantum resources to yield higher statistical precision of unknown
parameters than purely classical approaches \cite{review}. It is of particular interest in
quantum metrology, quantum lithography, gravity-wave detection and quantum computation.
A lot of work has been done, both
theoretically \cite{t1,t2,t3,t4} and experimentally \cite{e1,e2,e3,e4}, to exploit the quantum
advantages over classical measurement strategy. Simultaneously estimating more than one
parameters represents an interesting possibility to extend the concept of quantum metrology.
One application of this new technique of quantum metrology to the wider research community
is microscopy. Recently, the quantum enhanced imaging making use of point estimation theory
is presented \cite{imaging}, and the vector phase estimation is then investigated since phase
imaging is inherently a vector parameter estimation problem \cite{mutiple}. With respect to
the mature experimental techniques of multiqubit manipulation and multi-port devices \cite{integrated},
there is an urgent demand for the theoretic study of the multi-mode quantum metrology. The
vector parameter estimation technique will also be of significant use for other science and
technology areas such as multipartite clock synchronization \cite{qcs} and gravity-wave
detection \cite{ligo}.

Humphreys \emph{et al.} \cite{mutiple} find that quantum simultaneous estimation (SE) strategy
provides an advantage in the total variance of a vector parameter over optimal individual estimation
(IE) schemes, which remarkably scales as $\mathcal{O}(d)$ with $d$ the number of parameters.
This result is obtained via quantum Cram\'{e}r-Rao bounds (QCRBs) corresponding to the quantum
Fisher information (QFI) matrix for unitary evolution in the absence of noise. However, the most
well known QCRBs are asymptotically tight in limit of infinitely many trials and can grossly
underestimate the achievable error when the likelihood function is highly non-Gaussian \cite{zzb,extendedzzb}.
The QCRBs are merely supposed to be obtained when the prior distribution of the parameter is
peaked around the value of the parameter. In consideration of the nontrivial prior information of
vector parameter, extension of quantum Ziv-Zakai bounds \cite{qzzb} (QZZBs) that relates mean
square error to the probability of error in binary detection problem seems a superior alternative
in quantum vector parameter estimation.

Here, we derive two kinds of quantum metrological bounds for vector parameter with
uniform prior distribution by extending QZZBs to the vector parameter case. These bounds invoking
the quantum speed limit theorem \cite{speedlimit} indicate that only a constant advantage
can be obtained via SE strategy over the optimal IE strategy with NOON states. In addition to the
analysis of optimal probe states, the SE scheme with multimode squeezed vacuum states as a
promising optical resource has also been investigated and compared to IE with two-mode squeezed
vacuum states. Due to the fact that all realistic experiments face the decoherence, a general
framework for obtaining the lower bounds in the noisy system are also proposed. The lower bounds
are applied to evaluate the precision under two important decoherence models: photon
loss model and phase diffusion model \cite{phasediffusion}.

\begin{figure}[b]
 \centering
\includegraphics[width=0.45\textwidth]{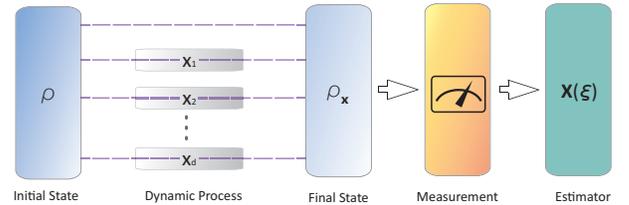}\\
\caption{{Vector parameter estimation procedure.}  $d$ independent dynamic maps corresponding
to $d$ parameters $\bm{x}$ transform the initial state $\rho$ to the final state $\rho_{\bm{x}}$. The
vector estimator $\bm{X}(\xi)$ is obtained according to results $\bm{\xi}$ from the measurements
performed on the final state $\rho_{\bm{x}}$.}\label{fig1}
\end{figure}
\section{Quantum metrological bounds for vector parameter estimation}
In the vector parameter estimation procedure, let $\bm{x}=(x_{1},\cdots,x_{m})^{T}$ be a
$d$-dimensional continuous vector random parameter and let $\bm{x}$ have a {prior} probability
density function ({PDF}) $P_{\bm{x}}(\bm{x})$. We may obtain finite measurement results as
$\bm{\xi}=(\xi_{1},\xi_{2},\cdots)^{T}$ to calculate the vector estimator
$\bm{X}(\bm{\xi})=(X_{1}(\bm{\xi}),\cdots,X_{d}(\bm{\xi}))^{T}$ with the observation conditional {PDF}
$P_{\bm{\xi}|\bm{x}}(\bm{\xi}|\bm{x})$ of $\bm{\xi}$ given the true values $\bm{x}$. This conditional
{PDF} plays a centre role in the investigation of QFI and QCRB.

Instead of the covariance matrix
$\bm{\Sigma}_{\bm{x}}$ discussed in Ref.~\cite{mutiple}, we pay our attention to the estimation error,
$\bm{\epsilon}=\bm{X}(\bm{\xi})-\bm{x}$. The mean-square estimation error correlation matrix is
defined as \cite{extendedzzb}
$\bm{\Sigma}_{\bm{\epsilon}}=\int d\bm{x}d\bm{\xi}P_{\bm{x},\bm{\xi}}(\bm{x},\bm{\xi})\bm{\epsilon}\bm{\epsilon}^{T}$,
where $P_{\bm{x},\bm{\xi}}(\bm{x},\bm{\xi})=P_{\bm{\xi}|\bm{x}}(\bm{\xi}|\bm{x})P_{\bm{x}}(\bm{x})$
denotes the joint {PDF} of $\bm{x}$ and $\bm{\xi}$. The total mean square error is obtained
by taking a trace of this error correlation matrix
\begin{equation}
|\bm{\Sigma}_{\bm{\epsilon}}| \equiv \textrm{Tr}\left(\bm{\Sigma}_{\bm{\epsilon}}\right)
=\sum_{i=1}^{d}\int d\bm{x}d\bm{\xi}P_{\bm{x},\bm{\xi}}(\bm{x},\bm{\xi})[{X}_{i}(\bm{\xi})-{x}_{i}]^{2},
\end{equation}
with which we will be concerned in the rest of our paper. Ziv-Zakai bounds \cite{zzb} (ZZBs)
assume that the parameter is a random variable with a known prior PDF, while
QCRBs treat the parameter as an unknown deterministic vector of quantities. Therefore, QZZBs \cite{qzzb}
that relate the mean-square error to the error probability in a binary hypothesis testing problem are
superior alternatives for obtaining the lower bounds when taking the prior information into consideration.
In this Letter, let us assume for the moment that the prior distribution of the vector parameter is a uniform window with
mean $\bm{\mu}=(\mu_{1},\cdots,\mu_{d})^{T}$ and width $\bm{W}=(W_{1},\cdots,W_{d})^{T}$:
\begin{eqnarray}
P_{\bm{x}}(\bm{x})=\prod_{i=1}^{d}\frac{1}{W_{i}}\textrm{rect}\left(\frac{x_{i}-\mu_{i}}{W_{i}}\right),
\label{uniform}
\end{eqnarray}
which means that we have no prior information on the vector parameter before the estimate. This
assumption is reasonable because it has been demonstrated that in the high prior information
regime, the resulting accuracy is of same order one obtainable by guessing a random value in
accordance to the prior PDF \cite{zzeb}.

As shown in Fig.~\ref{fig1}, general vector parameter estimation procedure can be divided into three
distinct sections: probe preparations, interaction between the probe and the system, and the
probe readouts for determining estimators \cite{review}. Consider that the interaction between
the probe states and the system with the unknown vector parameter can be expressed as a
unitary operator $\rho_{\bm{x}}=\exp(-i\bm{H}^{T}\bm{x})\rho\exp(i\bm{H}^{T}\bm{x})$
where $\bm{H}=(\hat{H}_{1},\cdots,\hat{H}_{d})^{T}$ is a vector of Hamiltonians,
$[\hat{H}_{i},\hat{H}_{j}]=0$ and the initial state is a pure state $\rho=|\psi\rangle\langle\psi|$. In
this case,
the extended QZZBs for vector  parameter estimation can be written as the sum of
the lower bound for each parameter corresponding to a Hamiltonian (See Appendix for details.):
\begin{equation}
|\bm{\Sigma}_{\bm{\epsilon}}|
\geq\sum_{i=1}^{d}\int_{0}^{W_{i}}d\tau_{i}\frac{\tau_{i}}{2}\left(1-\frac{\tau_{i}}{W_{i}}\right)
\left(1-\sqrt{1-F_{i}^{2}(\tau_{i})}\right),
\label{ea}
\end{equation}
where $F_{i}(\tau_{i})\equiv|\langle\psi|\exp(-i\hat{H}_{i}\tau_{i})|\psi\rangle|$ denotes to the
fidelity of a single parameter.

Then the problem is translated to evaluation of the fidelity of a single
parameter  which is a centre and widely studied problem in quantum speed
limit theorem \cite{speedlimit,openspeedlimit} and quantum metrology \cite{qzzb}.
Thus, it is possible to obtain
two kinds of lower bounds using two different approximations \cite{qzzb}:
$F^{2}_{i}(\tau_{i})\geq1-2\lambda\tau_{i}\langle\hat{H}_{i}\rangle_{+}$ and
$F^{2}_{i}(\tau_{i})\geq\cos^{2}\Delta\hat{H}_{i}\tau_{i}$, where
$\langle\hat{H}_{i}\rangle_{+}\equiv\langle\psi|\hat{H}_{i}|\psi\rangle-E_{i}$
with $E_{i}$ the minimum eigenvalue of $\hat{H}_{i}$, $\lambda\simeq0.7246$ and  $\Delta\hat{H}_{i}^{2}\equiv\langle\psi|\hat{H}_{i}^{2}|\psi\rangle-\langle\psi|\hat{H}_{i}|\psi\rangle^{2}$.
Hence, it is possible to obtain a Margolus-Levitin (ML) type bound depending on the effective average
energy \cite{MLB} $|\bm{\Sigma}_{\bm{\epsilon}}|\gtrsim\sum_{i=1}^{d}\frac{1}{80\lambda^{2}\langle\hat{H}_{i}\rangle_{+}^{2}}$
for $W_{i}\gg{1}/({2\lambda\langle\hat{H}_{i}\rangle_{+}})$ and a Mandelstam-Tamm (MT) type bound
limited by the fluctuation of the Hamiltonian \cite{MTB}:
$|\bm{\Sigma}_{\bm{\epsilon}}|\gtrsim\sum_{i=1}^{d}\frac{\pi^{2}/16-1/2}{\Delta\hat{H}_{i}^{2}}$ for
$W_{i}\gg{\pi}/({2\Delta\hat{H}_{i}})$.
The MT type bound is capable of predicting the same scaling with QCRBs of multiple parameter
estimation \cite{mutiple}, however, may be less tight than QCRBs for some cases. The ML type
bound scaling with the average energy relative to the ground state is a new bound for multiple
parameter estimation. Moreover, we will show that the ML type bound exists even if the ground
energy of Hamiltonian is negative infinite or energy eigenvalues are continuous. It is necessary to
emphasize that both bounds, ML type or MT type, include the prior information and should be
satisfied simultaneously.
Therefore,  when $W_{i}\gg\max\{{1}/({2\lambda\langle\hat{H}_{i}\rangle_{+}}),{\pi}/({2\Delta\hat{H}_{i}})\}$
holds, the acceptable lower bound from QZZBs is written as
\begin{eqnarray}
|\bm{\Sigma}_{\bm{\epsilon}}|\gtrsim\max\left\{\sum_{i=1}^{d}\frac{c_{ML}}{\langle\hat{H}_{i}\rangle_{+}^{2}},\sum_{i=1}^{d}\frac{c_{MT}}{\Delta\hat{H}_{i}^{2}}\right\},
\label{zzbt}
 \end{eqnarray}
where we let $c_{ML}=1/80\lambda^{2}$ and $c_{MT}=\pi^{2}/16-1/2$. Letting $d=1$, these lower
bounds will reduce into the results discussed in Ref.~\cite{qzzb} which conform to the mainstream understanding
of the Heisenberg limit \cite{hl,t4}.

From Eq.~(\ref{zzbt}), the vector parameter estimation problem is decomposed into $d$ single
parameter estimation problems and the solution of each problem is affordable by the investigation on the
effective average and the variance of the Hamiltonian of each mode given the same state. This result is
of great importance because the conclusions from the study of single parameter estimation give direct
operational significance to the vector parameter estimation.

\begin{figure}[t]
 \centering
\includegraphics[width=0.3\textwidth]{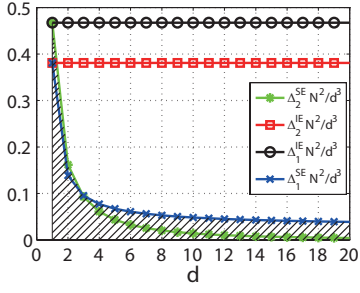}\\
\caption{{SE v.s. IE for optimal state.} Four bounds, multiplied by $N^{2}/d^{3}$, of optimal probe state:
ML bound $\Delta_{1}^{SE}$ and MT bound $\Delta_{2}^{SE}$ for SE strategy together with ML bound
$\Delta_{1}^{IE}$ and MT bound $\Delta_{2}^{IE}$ for IE strategy are compared against $d$. The hatched
area represents the forbidden value of precision for SE strategy.}
\label{fig2}
\end{figure}

\section{Optimal probe states and multimode squeezed states without noises}
Generally, the precision in determining the phase in interferometry is bounded by the inverse of the mean
total number of photons, which is the Heisenberg limit to optical interferometry. Recently, the validity of
this bound has been challenged by QCRBs under some situations \cite{dd1,dd2}, where the the QCRBs can
be even arbitrarily low. In these cases, the QZZBs in view of prior information on the parameter can be
much tighter and more effective than QCRBs.  Recently, it is derived from QCRBs that the advantage of
SE over the optimal IE is at best $\mathcal{O}(d)$ \cite{mutiple}. Here, motivated by the problem in the
single parameter estimation case, we will use QZZBs to investigate the performance of SE against IE when
taking the prior information of vector parameter into consideration.  We will try to find whether
$\mathcal{O}(d)$ advantage is effective given low and independent prior information in the vector
parameter case.

Hamiltonian operators are considered as $\hat{H}_{i}=\hat{n}_{i}$, and we label the optimal probe state
of totally $N$ particles in the Fock space as \cite{mutiple}
\begin{equation}
|\psi_{o}\rangle=\beta|N\underbrace{0\cdots0}_{d}\rangle+\alpha(|0N0\cdots0\rangle+\cdots+|0\cdots0N\rangle),
\label{states}
\end{equation}
with $d\alpha^{2}+\beta^{2}=1$ and $\alpha=1/\sqrt{d+\sqrt{d}}$. The effective average and the variance
for the $i$th Hermitian can be calculated as $\langle\hat{n}_{i}\rangle_{+}=\alpha^{2}N$ and
$\Delta\hat{n}_{i}^{2}=\alpha^{2}(1-\alpha^{2})N^{2}$ which lead to the lower bounds for SE
strategy as
\begin{eqnarray}
|\bm{\Sigma}_{\bm{\epsilon}}|\geq\max\left\{\frac{d(d+\sqrt{d})^{2}c_{ML}}{N^{2}},\frac{d(d+\sqrt{d})^{2}c_{MT}}{(d+\sqrt{d}-1)N^{2}}\right\}.
\label{optimalstate}
 \end{eqnarray}
 Here the first bound is a ML type bound labelled as $\Delta_{1}^{SE}$, and the second one is a MT type lower
bound labelled as $\Delta_{2}^{SE}$. The best quantum strategy of IE uses NOON states with total $N/d$
photons for each parameter. Thus, the ML bound for IE is
$\Delta_{1}^{IE}\simeq{d^{3}}/({20\lambda^{2}N^{2}})$ and the MT for IE is
$\Delta_{2}^{IE}\simeq{d^{3}(\pi^{2}-8)}/({4N^{2}})$.

The combined bounds for SE strategy and IE
strategy uniformly multiplied by $N^{2}/d^{3}$ are compared in Fig.~\ref{fig2}. For IE strategy, it is easy to conclude that
MT type bound $\Delta_{2}^{IE}$ is tighter. For SE strategy, the hatched area represents the forbidden
value due to the combined bound (\ref{optimalstate}). For large value of $d$, ML bound is
tighter, while MT bound is tighter for small value of $d$. Although for large $d$, the MT bound for SE
presents the $\mathcal{O}(d)$ advantage over both types of bounds for IE strategy, the ML bound that
should also be fulfilled denies this advantage and merely a constant advantage
($\lim_{d\rightarrow\infty}\Delta_{2}^{IE}/\Delta_{1}^{SE}\simeq4.9081$) can be observed.
Generally, it is possible to demonstrate that for any multiparticle entangled state with $N$ total particle
number, no $\mathcal{O}(d)$ advantage can be gained via SE strategy over IE strategy.
That is to say, QCRBs still underestimate the measurement precision in the vector parameter estimation case,
which should also be ascribed to not considering the prior information. The $\mathcal{O}(d)$ advantages
predicted by QCRBs are not effective in the vector parameter estimation.

Another entangled state we focus on is the $(d+1)$-mode squeezed vacuum
state \cite{multimodesqeezedstate1,multimodesqeezedstate2} which may not provide
a more accurate measurement of vector parameter but is more promising in the real application
with existing experimental techniques. Recently, it has been reported that advanced LIGO
can be improved with squeezed vacuum states \cite{ligo}.
Whether $(d+1)$-mode squeezed vacuum state can provide advantage over the IE strategy via using
two-mode squeezed vacuum state to estimate each parameter individually is an interesting and
important problem.

\begin{figure}[t]
 \centering
\includegraphics[width=0.47\textwidth]{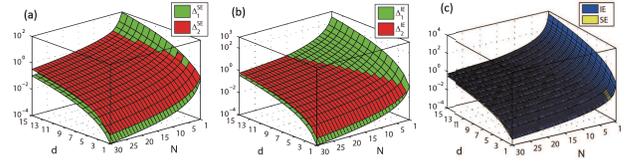}\\
\caption{{SE v.s. IE for multimode squeezed vacuum state.} ({a}) ML bound $\Delta_{1}^{SE}$ and MT bound $\Delta_{2}^{SE}$ for SE strategy using the $(d+1)$-mode squeezed vacuum state against $d$ and total average particle number $N$.
({b}) ML bound $\Delta_{1}^{IE}$ and MT bound $\Delta_{2}^{IE}$ for IE strategy using $d$ two-mode squeezed vacuum states against $d$ and total average particle number $N$.
({c}) The combined bounds for SE strategy and the one for IE strategy are compared against $d$ and total average particle number $N$.}
\label{fig3}
\end{figure}

Using the relation between the Bose operators ($\hat{a}_{i}$, $\hat{a}_{i}^{\dag}$), the coordinate operator
$\hat{Q}_{i}=(\hat{a}_{i}+\hat{a}_{i}^{\dag})/\sqrt{2}$ and the momentum operator
$\hat{P}_{i}=(\hat{a}_{i}-\hat{a}_{i}^{\dag})/(i\sqrt{2})$, one can express the $(d+1)$-mode
squeezed operator $\hat{S}(r)$ with form \cite{multimodesqeezedstate2}
$\hat{S}(r)=\exp\left(ir{\bm{Q}^{T}}{A}\bm{P}\right)$
where $\bm{Q}=(\hat{Q}_{0},\hat{Q}_{1},\cdots,\hat{Q}_{d})^{T}$, $\bm{P}=(\hat{P}_{0},\hat{P}_{1},\cdots,\hat{P}_{d})^{T}$ and
${A}$ is a $(d+1)\times(d+1)$ matrix with elements
${A}_{kj}=\delta_{k,(j+1)\textrm{mod}(d+1)}$ and $k,j=0,1,\cdots,d$.

Via the Campbell-Baker-Hausdorff formula, we find that the squeezing operator
transforms the annihilation operators as
$\hat{S}^{\dag}(r)\hat{a}_{k}\hat{S}(r)
=\sum_{i=0}^{d}\left({R_{ki}\hat{a}_{i}+K_{ki}\hat{a}^{\dag}_{i}}\right)$ \cite{multimodesqeezedstate2}
with $R=(e^{-r\tilde{A}}+e^{r{A}})/2$, $K=(e^{-r\tilde{A}}-e^{r{A}})/2$ and $\tilde{A}$ is the transpose of $A$.
Because we have that $\tilde{A}^{D}=A^{D}=\mathbb{I}$ with $D=d+1$,
it can be expanded as
$\exp(-r\tilde{A})=c_{0}\mathbb{I}+c_{1}\tilde{A}+c_{2}\tilde{A}^{2}+\cdots+c_{d}(r)\tilde{A}^{d}$ with
$\tilde{A}_{kj}^{m}=\delta_{(k+m)\textrm{mod}D,j}$ and ${A}_{kj}^{m}=\delta_{k,(j+m)\textrm{mod}D}$, and the coefficients are given by the following equation:
\begin{equation}
\left(\begin{array}{c}
c_{0}\\
c_{1}\\
\vdots\\
c_{d}
\end{array}\right)
=\frac{1}{D}
\left(\begin{array}{c c c c c}
1&1&
\cdots&1\\
1&e^{-i\omega_{D}}&
\cdots&e^{-id\omega_{D}}\\
\vdots&\vdots&
\ddots&\vdots\\
1&e^{-id\omega_{D}}&
\cdots&e^{-id^{2}\omega_{D}}
\end{array}\right)
\left(\begin{array}{c}
e^{-r}\\
e^{-re^{i\omega_{D}}}\\
\vdots\\
e^{-re^{id\omega_{D}}}
\end{array}\right),
\end{equation}
where $\omega_{D}=2\pi/D$. The photon number operator for the $k$-th mode is $
\hat{n}_{k}=\hat{a}^{\dag}_{k}\hat{a}_{k}$.
The average of the photon number operator is
$\langle \hat{n}_{k}\rangle
=\frac{1}{4}\sum_{i=0}^{d}\left(c_{i}^{2}+c_{i}^{2}\right)-\frac{1}{2}$,
where the average is taken on $|\bm{0}\rangle=\bigotimes_{i=0}^{d}|0_{i}\rangle$, and we have used the fact that $\sum_{i}(e^{-r\tilde{A}})_{ki}(e^{r{A}})_{ki}=\sum_{i}(e^{-r\tilde{A}})_{ki}(e^{r\tilde{A}})_{ik}=1$.
The variance of the photon number operator is calculated as
\begin{eqnarray}
(\Delta \hat{n}_{k})^{2}
=\frac{1}{8}\left(\sum_{i=0}^{d}c_{i}^{2}\right)^{2}+\frac{1}{8}\left(\sum_{i=0}^{d}c_{i}^{2}\right)^{2}-\frac{1}{4}.
\end{eqnarray}
Here, we numerically evaluate the performance of multi-mode squeezed state in
the multiple parameter estimation procedure.

Our SE strategy using the $(d+1)$-mode squeezed vacuum state $\hat{S}(r)|{0\cdots0}\rangle$
is compared with the IE strategy using two-mode squeezed vacuum state. The total average
particle numbers for the  $(d+1)$-mode squeezed vacuum state and $d$ identical two-mode
squeezed vacuum states are set the same. Numerical results of lower bounds for SE and IE are
presented in Fig.~\ref{fig3}(a), (b), and they are compared in Fig.~\ref{fig3}(c). These results indicate
that the SE strategy with $(d+1)$-mode squeezed vacuum state is superior to the IE strategy
except for the circumstances where $d=2$ and the value of total average photon number $N$
is around $2$.

\begin{figure*}[t]
 \centering
\includegraphics[width=0.85\textwidth]{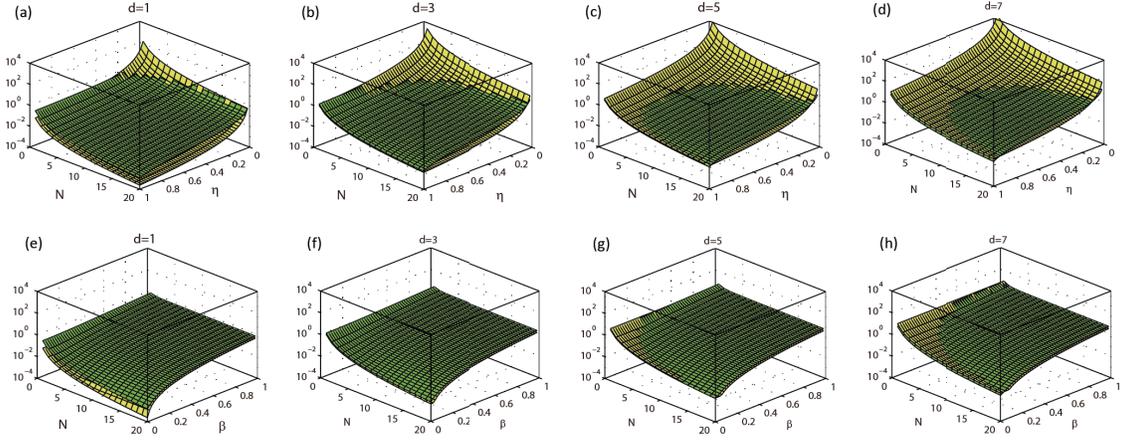}\\
\caption{{Noisy ML bound v.s. MT bound for optimal probe state.}
Green curved surface is for $\Delta_{2}^{SE}$ and the yellow one is for $\Delta_{1}^{SE}$.
(a)-(d) are for the photon loss model: two bounds for state (\ref{states}) are compared against particle number $N$ and intensity of the transmissivity $\eta$. (e)-(h) are for the phase diffusion model: two bounds for state (\ref{states}) are compared against particle number $N$ and diffusion parameter $\eta$.}
\label{fig4}
\end{figure*}

\section{Quantum metrological bounds in noisy systems}
The quantum metrological bounds in noisy systems have become a focus of attentions because
in real experiments there will always be some degree of noise and limitation.
For unitary processes, the analytical expressions of the lower bound for estimating multiple parameter
have been established, however, for noisy case, the task may be of exceptional difficulty since saturation
of QFI matrix for vector parameters is, in general, difficult or impossible.

Here we generalize our analysis to vector parameter estimation in
noisy system. Since the noisy dynamic process is assumed to include $d$ independent quantum channels
corresponding to $d$ parameters, respectively, it can be described as
$\rho_{\bm{x}}=\sum_{\bm{l}}\hat{\Pi}_{\bm{l}}(\bm{x})\rho\hat{\Pi}_{\bm{l}}^{\dag}(\bm{x})$ in terms of
Kraus operators $\{\hat{\Pi}_{\bm{l}}(\bm{x})=\bigotimes_{i=1}^{d}\hat{\pi}_{l_{i}}^{(i)}(x_{i})\}$ with
$\bm{l}=(l_{1},l_{2},\cdots,l_{d})$ under
identity condition $\sum_{l_{i}}\hat{\pi}_{l_{i}}^{(i)\dagger}(x_{i})\hat{\pi}_{l_{i}}^{(i)}(x_{i}) = \mathbb{I}$.
This dynamic process has another equivalent description expressed as tracing environment after a unitary evolution operators
$U_{SE}(\bm{x})=\bigotimes_{i=1}^{d}U^{(i)}_{SE}({x}_{i})$ acting on the pure state $|\psi\rangle|\bm{0}\rangle_{E}$ of an
enlarged space for the system interacting with an environment. With these preconditions and Uhlmann's theorem,
the fidelity of two states is bounded as
\begin{eqnarray}
&&F^{2}(\rho_{\bm{x}},\rho_{\bm{x}+\bm{\tau}_{i}})\nonumber\\
&\geq&\left|{}_{E}\langle\bm{0}|\langle\psi|U^{\dag}_{SE}(\bm{x})U_{SE}(\bm{x}+\bm{\tau}_{i})|\psi\rangle|\bm{0}\rangle_{E}\right|^{2}\nonumber\\
&=&\left|\langle\psi|\sum_{{l}_{i}}\hat{\pi}_{{l}_{i}}^{\dag}({x}_{i})\hat{\pi}_{{l}_{i}}({x}_{i}+{\tau}_{i})|\psi\rangle\right|^{2}\equiv\mathcal{F}^{2}_{i}({\tau}_{i}),
\label{ggg}
\end{eqnarray}
where $|\bm{0}\rangle_{E}=\bigotimes_{i=0}^{d}|0_{i}\rangle_{E}$ and we have assumed that $\mathcal{F}_{i}({\tau}_{i})$, the
fidelity of the enlarged system, only corresponds to the difference of the parameter. 

Here, let the unitary matrix $\bigotimes_{i=1}^{d}{u}_{E}^{(i)}(x_{i})$ relate different purifications of the
final states $\rho_{\bm{x}}$ and connect different sets of linearly independent Kraus operators \cite{elusive},
as well.
Then, the Hermitian generator for $i$th mode should be written as \cite{escher2012}
\begin{equation}
\hat{\mathcal{H}}_{i}(x_{i})\equiv-i\frac{d[u_{E}^{(i)}(x_{i})U_{SE}^{(i)}(x_{i})]^{\dag}}{dx_{i}}u_{E}^{(i)}(x_{i})U_{SE}^{(i)}(x_{i}). \label{operator}
\end{equation}
For a parameter-independent Hermitian generator $\hat{\mathcal{H}}_{i}$ with
$\mathcal{F}^{2}({\tau}_{i})=|{}_{E}\langle\bm{0}|\langle\psi|\exp(-\hat{\mathcal{H}}_{i}x_{i})|\psi\rangle|\bm{0}\rangle_{E}|^{2}$,
one is able to obtain the rational quantum metrology bounds limited by the bounds
of the unitary process of enlarged system. Given the uniform prior distribution
(\ref{uniform}), the lower bound Eq.~(\ref{a1}) in APPENDIX can be lower bounded by the
fidelity of the enlarged system as
\begin{eqnarray}
|\bm{\Sigma}_{\bm{\epsilon}}|&\geq&\sum_{i=1}^{d}\int_{0}^{W_{i}}d\tau_{i}\frac{\tau_{i}}{2}\left(1-\frac{\tau_{i}}{W_{i}}\right)\left(1-\sqrt{1-\mathcal{F}^{2}({\tau}_{i})}\right)\nonumber\\
&\geq&\max\left\{\sum_{i=1}^{d}\frac{c_{ML}}{\langle\hat{\mathcal{H}}_{i}\rangle_{+}^{2}},\sum_{i=1}^{d}\frac{c_{MT}}{\Delta\hat{\mathcal{H}}_{i}^{2}}\right\}.
\end{eqnarray}
To make the lower bounds tighter we should consider all the possible forms of the
effective Hermitian generators. Tighter bounds can be written as
\begin{eqnarray}
|\bm{\Sigma}_{\bm{\epsilon}}|\geq\max\left\{\sum_{i=1}^{d}\frac{c_{ML}}{\min\langle\hat{\mathcal{H}}_{i}\rangle_{+}^{2}},\sum_{i=1}^{d}\frac{c_{MT}}{\min\Delta\hat{\mathcal{H}}_{i}^{2}}\right\}.
\label{noise}
\end{eqnarray}
Here the average is taken on state $|\psi\rangle|\bm{0}\rangle_{E}$, and the minimum runs over all
the possible forms of the unitary operator $u_{E}^{(i)}(x_{i})$ acting on the environment of $i$th mode.
 For MT bound, the minimum variance of $\hat{\mathcal{H}}_{i}$ multiplied by $4$ equals to the
attainable QFI of $i$th parameter alone in noisy system and has been discussed
in Refs.~\cite{escher2011general,escher2012}. ML bound in presence of the noise that corresponds to the
effective average of $\hat{\mathcal{H}}_{i}$ is a new result even for the single parameter estimation.
It concerns with the residual resource for estimating the parameter after suffering the noise. In addition,
we should  point out that the optimal unitary matrix that minimizes ML bound need
not be the same as the optimal one for MT bound.

To demonstrate the practicability of these lower bounds, we consider two common but
significant models in the noisy optical interferometry: photon loss model and phase diffusion model.
We will show that both bounds, MT type or ML type,  need to be satisfied when taking these
noise models into account. In particular, our lower bounds that include the prior information of vector
parameter are shown to be versatile in different quantum metrology problems and much
tighter than the popular QCRBs in certain cases.

\subsection{Photon loss model}
In some noisy models, the physical processes are described by the Krauss operators, and the analytical
expression of effective Hermitian generator for enlarged system may not be easily obtained. The photon
loss model is such a model. However, this problem can be somewhat settled by similar calculations we
have done in the study of quantum speed limit.

A possible set of Kraus operators describing the photon loss model in $i$th mode is written as \cite{escher2011general}:
\begin{equation}
\hat{\pi}_{l_i}^{(i)} = \sqrt{\frac{(1-\eta_{i})^{l_i}}{l_i !}}e^{i x_{i} (\hat{n}_i - \delta_i l_i)} \eta_i^{\frac{\hat{n}_i}{2}}\hat{a}_{i}^{l_i},\ i=1,\cdots,d,
\end{equation}
given $\hat{n}_{i}=\hat{a}_{i}^{\dag}\hat{a}_{i}$ the number operator on $i$th mode, $\eta_{i}$ the
intensity of the transmissivity and $\delta_{i}$ the variational parameter.
From Eq.~(\ref{ggg}), we obtain for the optimal probe state (\ref{states}) that
\begin{eqnarray}
\mathcal{F}^{2}_{i}({\tau}_{i})&=&\left|\langle\psi_{o}|\sum_{{l}_{i}}\frac{(1-\eta_{i})^{l_{i}}}{l_{i}!}(\hat{a}^{\dag})^{l_{i}}\eta_{i}^{\hat{n}_{i}}e^{i\tau_{i}(\hat{n}_{i}-\delta_{i}l_{i})}\hat{a}^{l_{i}}|\psi_{o}\rangle\right|^{2}\nonumber\\
&=&\left|\sum_{l_{i}=0}^{N+1}P_{l_{i}}e^{i\tau_{i}E_{l_{i}}}\right|^{2}
\end{eqnarray}
where $E_{(N+1)_{i}}=0$ and $E_{l_{i}}=N-\sigma_{i}l_{i}$ with probabilities $P_{(N+1)_{i}}=1-\alpha^{2}$ and
$P_{l_{i}}=\alpha^{2}\left(_{l_{i}}^{N}\right)(1-\eta_{i})^{l_{i}}\eta_{i}^{N-l_{i}}$, and we let $\sigma_{i}=\delta_{i}+1$.
Following the calculations performed in the quantum speed limit to dynamical evolution \cite{speedlimit},
we can construct a effective Hermitian operator $\hat{\mathcal{H}}_{i}^{pl}$ with  its effective average
$\langle\hat{\mathcal{H}}_{i}^{pl}\rangle_{+}=\langle\hat{n}_{i}\rangle[1-\sigma_{i}(1-\eta_{i})]-\min\{N(1-\sigma_{i}),0\}$ and its variance
$\Delta(\hat{\mathcal{H}}_{i}^{pl})^{2}
=\Delta\hat{n}_{i}^{2}[1-\sigma_{i}(1-\eta_{i})]^{2}+\langle\hat{n}_{i}\rangle\sigma_{i}^{2}\eta_{i}(1-\eta_{i})$,
where $\langle\hat{n}_{i}\rangle=\alpha^{2}N$, $\Delta\hat{n}_{i}^{2}=\alpha^{2}(1-\alpha^{2})N^{2}$ and
$\min\{N(1-\sigma_{i}),0\}$ denotes the ground energy eigenvalue.
The minimums of these two quantities for the optimal variational parameter contribute to the lower bound
for photon loss channel:
\begin{equation}
|\bm{\Sigma}_{\bm{\epsilon}}|^{pl}\geq\max\left\{\sum_{i=1}^{d}\frac{c_{ML}}{\eta_{i}^{2}\langle\hat{n}_{i}\rangle^{2}},\sum_{i=1}^{d}\frac{c_{MT}}{\frac{\eta_{i}\langle\hat{n}_{i}\rangle\Delta\hat{n}_{i}^{2}}{(1-\eta_{i})\Delta\hat{n}_{i}^{2}+\eta_{i}\langle\hat{n}_{i}\rangle}}\right\},
\end{equation}
where $\delta_{i}^{opt}=1$ is the optimal variational parameter for ML type bound, and $\delta_{i}^{opt'}=\frac{\Delta\hat{n}_{i}^{2}}{(1-\eta_{i})\Delta\hat{n}_{i}^{2}+\eta_{i}\langle\hat{n}_{i}\rangle}-1$
is for MT type bound.

For simplicity, we suppose that
different modes of lossy channels share the same lossy parameter:  the intensity of the
transmissivity of $i$th mode has $\eta_{i}=\eta$. For different values of $d$, the ML bounds (yellow
curved surface) and MT bounds (green curved surface), given state (\ref{states}), are compared in Fig.~\ref{fig4}(a)-(d).
It is clearly
displayed that under some conditions MT bound are tighter than ML bound and vice versa. It
shows that both bounds MT type or ML type are essential and need to be satisfied when taking
the effect of photon loss into account.

\subsection{Phase diffusion model}
The phase diffusion noise can be represented, in the Markov limit, by the unitary for system and
environment under the situation that $\sqrt{2}\beta_{i}^{2}n_{i}\gg1$ \cite{escher2012}:
\begin{eqnarray}
&&u_{E}(\bm{x})U_{SE}(\bm{x})=\bigotimes_{i=1}^{d}e^{-ix_{i}\hat{n}_{i}}e^{ix_{i}\kappa_{i}\hat{P}^{E}_{i}/2\beta_{i}}e^{i2\beta_{i}\hat{n}_{i}\hat{Q}_{i}^{E}}\nonumber\\
&=&\bigotimes_{i=1}^{d}\left[\sum_{n_{i}}|n_{i}\rangle\langle n_{i}|e^{-ix_{i} n_{i}}\mathcal{D}_{E}\left(\alpha_{1}+\alpha_{2}\right)e^{\frac{ix_{i}\kappa_{i} n_{i}}{2}}\right],
\end{eqnarray}
where $\alpha_{1}=-\frac{x_{i}\kappa_{i}}{2\sqrt{2}\beta_{i}}$, $\alpha_{2}=i\sqrt{2}\beta_{i} n_{i}$,  $\hat{Q}^{E}_{i}=\frac{\hat{b}_{i}+\hat{b}_{i}^{\dag}}{\sqrt{2}}$ and $\hat{P}^{E}_{i}=\frac{\hat{b}_{i}-\hat{b}_{i}^{\dag}}{i\sqrt{2}}$ with
$\hat{b_{i}}$ and $\hat{b}_{i}^{\dag}$ the
annihilation and creation operators acting on the environment corresponding to the $i$th mode of the
system. For the $i$th mode, $\beta_{i}$ stands for the diffusion parameter and $\kappa_{i}$ is the variational
parameter. $\mathcal{D}_{E}(\alpha)$ is the displacement operator on the environment.
Here, we have used the fact that
$\mathcal{D}_{E}(\alpha_{1})\mathcal{D}_{E}(\alpha_{2})=\mathcal{D}_{E}(\alpha_{1}+\alpha_{2})\exp(i\textrm{Im}\{\alpha_{1}\alpha_{2}^{*}\})$.
The reduced state is
$\rho_{\bm{x}}=\textrm{Tr}_{E}(u_{E}(\bm{x})U_{SE}(\bm{x})|\psi\rangle|\bm{0}\rangle_{E}\langle\bm{0}|\langle\psi|U^{\dag}_{SE}(\bm{x})u_{E}^{\dag}(\bm{x}))$
where the state for enlarged system is
$|\psi\rangle|\bm{0}\rangle_{E}=\sum_{\bm{n}}\sqrt{P_{\bm{n}}}|\bm{n}\rangle|\bm{0}\rangle_{E}=\sum_{n_{0}\cdots n_{d}}\sqrt{P_{n_{0}\cdots n_{d}}}|n_{0}\cdots n_{d}\rangle|\bm{0}\rangle_{E}$.

In this model, the effective Hermitian operator for the enlarged system is obtained from Eq.~(\ref{operator})
as
$\hat{\mathcal{H}}_{i}^{pd}=(1-\kappa_{i})\hat{n}_{i}-{\kappa_{i}\hat{P}_{i}^{E}}/({2\beta_{i}})$,
and the eigenstate of this Hermitian operator is
$\bigotimes_{i,j=0}^{d}|n_{i}\rangle|p_{j}\rangle_{E}$
where $p_{j}\in(-\infty,\infty)$. Therefore, the calculation of the effective average may be impossible
because the eigenvalues of $\hat{\mathcal{H}}_{i}^{pd}$ are continuous and the ground energy eigenvalue
is negative infinity. In order to obtain the appropriate and nontrivial form of $\langle\hat{\mathcal{H}}\rangle_{+}$,
we restart from the form of the fidelity and obtain another version of the effective average which may
lead to a less tight bound. However, this effective average contains a clear physical implication. From
the expression of the fidelity, we have
\begin{widetext}
\begin{eqnarray}
&&F^{2}(\rho_{\bm{x}},\rho_{\bm{x}+\bm{\tau}_{i}})
\geq
\left|\sum_{\bm{n}}\int_{-\infty}^{\infty}dp_{i}P_{\bm{n}}|\langle\bm{n}|\psi\rangle|^{2}\frac{1}{\sqrt{\pi}}e^{-{p_{i}^{2}}}e^{-i\tau_{i}\left[(1-\kappa_{i})n_{i}-\frac{\kappa_{i}p_{i}}{2\beta_{i}}\right]}\right|^{2}\\
&\geq&1-\lambda\tau_{i}\sum_{n_{i}m_{i}}\int_{-\infty}^{\infty}dp_{i}dq_{i}\frac{|c_{n_{i}}c_{m_{i}}|^{2}}{{\pi}}e^{-{p_{i}^{2}-q_{i}^{2}}}\left|(1-\kappa_{i})(n_{i}-m_{i})-\frac{\kappa_{i}(p_{i}-q_{i})}{2\beta_{i}}\right|\\
&\geq&1-\lambda\tau_{i}\sum_{n_{i}m_{i}}|c_{n_{i}}c_{m_{i}}|^{2}\left|(1-\kappa_{i})(n_{i}-m_{i})\right|+\int_{-\infty}^{\infty}dp_{i}dq_{i}\frac{1}{{\pi}}e^{-{p_{i}^{2}-q_{i}^{2}}}\left|\frac{\kappa_{i}(p_{i}-q_{i})}{2\beta_{i}}\right|\geq1-2\lambda\tau_{i}\langle\hat{\mathcal{H}}_{i}^{pd}\rangle_{+},
\end{eqnarray}
\end{widetext}
where $|c_{n_{i}}|^{2}=\sum_{n_{0},\cdots,n_{i-1},n_{i+1},\cdots,n_{d}}P_{\bm{n}}|\langle \bm{n}|\psi\rangle|^{2}$, and
the effective average is obtained as
\begin{eqnarray}
\langle\hat{\mathcal{H}}_{i}^{pd}\rangle_{+}
=|1-\kappa_{i}|\langle\hat{n}_{i}\rangle+\frac{|\kappa_{i}|}{2\sqrt{2\pi}\beta_{i}}.
\end{eqnarray}
Given the values of $\beta_{i}$ and $\langle\hat{n}_{i}\rangle$, the tighter bound is obtained by the
optimal $\kappa_{i}$ that makes $\langle\hat{\mathcal{H}}_{i}^{pd}\rangle_{+}$ minimum.
The ML bound that contains a optimal problem is studied by numerical simulations and shown in Fig.~4(e)-(h).

For the MT bound, it is easy to obtain that
$\langle\hat{\mathcal{H}}_{i}^{pd}\rangle
=(1-\kappa_{i})\langle\hat{n}_{i}\rangle$ and
$(\Delta\hat{\mathcal{H}}_{i}^{pd})^{2}
=\Delta\hat{n}_{i}^{2}(1-\kappa_{i})^{2}+\frac{\kappa_{i}^{2}}{8\beta_{i}^{2}}$ \cite{escher2012}.
Considering that all the diffusion parameters are the same $\beta_{i}=\beta$ for $i=1,\cdots,d$,
one obtains the tighter bounds with
$\kappa_{i}=8\Delta n_{i}^{2}\beta^{2}/(1+8\Delta n_{i}^{2}\beta^{2})$.
For different values
of $d$, the ML bounds (yellow curved surface) and MT bounds (green curved surface), given state
(\ref{states}), are compared in Fig.~\ref{fig4}(e)-(h). It is clearly
displayed that under some conditions MT bound are tighter than ML bound and vice versa, as well.

\section{Conclusions and Discussions}
In this paper, we extend QZZBs to the multiple parameter case and present two kinds of lower bounds
in accordance with the quantum speed limit theorem. Compared with QCRBs lying on the infinitesimal
statistical distance between $\rho_{\bm{x}}$ and its neighborhood \cite{t1}, QZZBs depend on the
statistical distance between  $\rho_{\bm{x}}$ and $\rho_{\bm{x'}}$ for all relevant values of $\bm{x}$
and $\bm{x'}$. That is, our method provides a lower bound on the achievable precision in consideration
of the PDF characterizing the prior information of the vector parameter, which considers a more realistic
estimation problem. From quantum Ziv-Zakai error bounds, we have derived two kinds of quantum
metrological bounds for vector parameter estimation. We have shown that at best a constant advantage
can be expected via SE strategy over the optimal IE strateg. The SE scheme with multimode squeezed
vacuum states has also been investigated and compared to IE with two-mode squeezed vacuum states.
A general framework for obtaining the lower bounds in the noisy system has also been proposed with which we have
studied the photon loss model and phase diffusion model.

An important question not addressed above is the attainability of our twiformed lower bound. This
well-known saturation problem appears tough in two aspects. One is that, different from QCRBs in special cases,
the QZZBs are not expected to be satisfied and can not be used to study the optimal performance of
quantum parameter estimation \cite{zzb}. However, the QZZBs are shown to be much more versatile and tighter
than the popular QCRBs in many cases \cite{qzzb} such as single parameter estimation with squeezed states and
vector parameter estimation. Another aspect is that, the twiformed bound borrowing ideas from the
quantum limit theorem uses some approximations that may not be saturated. Nevertheless, our bounds
limited by both average and variance of the Hamiltonian present a much clearer physical meaning
which also includes that of QCRBs. A tighter bound can be obtained by the analytical or numerical studies
on the extended QZZBs in Eq.~(\ref{ea}).

Assessing the impact of noise on the performance of SE strategy for vector parameter estimation is
a crucial problem in quantum metrology and quantum imaging. It seems indeed difficult as the
attainability of QCRBs for noisy vector parameter estimation is not resolved or may be impossible.
Our investigation on the lower bounds in noisy system will perform a new tool for evaluating
noisy quantum metrology, even though it is not tight either. Moreover, unlike QCRBs, the results from
single parameter estimation will contribute directly to the case of vector parameter in noisy systems.

Since the QCRBs are more ``optimistic'' and supposed to be achieved only when the prior distribution
is peaked around the parameter, the attainability of QCRBs may be lost given a flat prior distribution.
However, in a realistic scenario for quantum parameter estimation, one can design schemes that increase
the prior information and set the observable with a better precision gradually during the measuring procedure.
For example, estimation strategies such as two-step adaptive schemes \cite{referee}  may be used in order
to achieve the QCRBs even though no prior information is known.

After the appearance of the first version of our manuscript, we notice that a related paper appears
to analyze the case that parameters have nontrivial prior correlations \cite{ntc}.

\begin{acknowledgments}
We would like to thank Augusto Smerzi and J.-D. Yue for useful discussions and suggestions.
This work was supported by the 973 Program (2010CB922904),
NSFC (11175248), grants from the Chinese Academy of Sciences.
\end{acknowledgments}

\appendix
\section{Quantum Ziv-Zakai bounds for vector parameter estimation}

Let $\epsilon_{i}\equiv|{X}_{i}(\bm{\xi})-{x}_{i}|$ be a nonnegative random variable.
The probability density of $\epsilon_{i}$ is the differentiation of cumulative probability:
$P_{\epsilon_{i}}(s_{i})=\frac{d}{ds_{i}}\Pr(\epsilon_{i}<s_{i})=-\frac{d}{ds_{i}}\Pr(\epsilon_{i}\geq s_{i})$.
Therefore, one obtains that
\begin{eqnarray}
|\bm{\Sigma}_{\bm{\epsilon}}|
=\sum_{i=1}^{d}\int_{0}^{\infty}d\tau_{i}\frac{\tau_{i}}{2}\Pr\left(|{X}_{i}(\bm{\xi})-{x}_{i}|\geq \frac{\tau_{i}}{2}\right),
\end{eqnarray}
where $\tau_{i}=2s_{i}$. Here, we can easily obtain that
\begin{widetext}
\begin{eqnarray}
&&\Pr\left(|{X}_{i}(\bm{\xi})-{x}_{i}|\geq \frac{\tau_{i}}{2}\right)
=\Pr\left({X}_{i}(\bm{\xi})\geq{x}_{i}+ \frac{\tau_{i}}{2}\right)
+\Pr\left({X}_{i}(\bm{\xi})\leq{x}_{i} -\frac{\tau_{i}}{2}\right)\\
&&=\int d\bm{\zeta}[P_{\bm{x}}(\bm{\zeta})+P_{\bm{x}}(\bm{\zeta}+\bm{\tau}_{i})]\left[P_{i}^{0}\Pr\left({X}_{i}(\bm{\xi})\geq{\zeta}_{i}+ \frac{\tau_{i}}{2}\Big|\bm{x}=\bm{\zeta}\right)
+P_{i}^{1}\Pr\left({X}_{i}(\bm{\xi})\leq{\zeta}_{i} +\frac{\tau_{i}}{2}\Big|\bm{x}=\bm{\zeta}+\bm{\tau}_{i}\right)\right],
\end{eqnarray}
\end{widetext}
where
$P_{i}^{0}=\frac{P_{\bm{x}}(\bm{\zeta})}{P_{\bm{x}}(\bm{\zeta})+P_{\bm{x}}(\bm{\zeta}+\bm{\tau}_{i})}$, $P_{i}^{1}=\frac{P_{\bm{x}}(\bm{\zeta}+\bm{\tau}_{i})}{P_{\bm{x}}(\bm{\zeta})+P_{\bm{x}}(\bm{\zeta}+\bm{\tau}_{i})}$,
$\bm{\zeta}=(\zeta_{1},\cdots,\zeta_{d})^{T}$ and  $\bm{\tau}_{i}=(\underbrace{0,\cdots,0}_{i-1},\tau_{i},0,\cdots,0)^{T}$.

\begin{table}[t]
\caption{\label{tab:table1} Binary hypothesis testing problem}
\begin{ruledtabular}
\begin{tabular}{c|c |c}
Two hypotheses&$\mathcal{H}_{i}^{0}:\ \bm{x}=\bm{\zeta}$&$\mathcal{H}_{i}^{1}:\ \bm{x}=\bm{\zeta}+\bm{\tau}_{i}$\\
\hline
Probability & ${\Pr}(\mathcal{H}_{i}^{0})=P_{i}^{0}$ & ${\Pr}(\mathcal{H}_{i}^{1})=P_{i}^{1}$\\
\hline
Estimation standard\footnote{The $i$th estimation standard (totally $d$ standards).} & $\delta_{i}^{0}:{X}_{i}(\bm{\xi})\leq \zeta_{i}+\frac{\tau_{i}}{2}$ & $\delta_{i}^{1}:{X}_{i}(\bm{\xi})\geq \zeta_{i}+\frac{\tau_{i}}{2}$\\
\end{tabular}
\end{ruledtabular}
\end{table}

Now let us consider a binary hypothesis testing problem with two hypotheses (see Table~\ref{tab:table1}).
The $i$th error probability (two kinds of error decision) of the binary hypothesis testing problem can be
written as \cite{qzzb}
\begin{eqnarray}
&&\textrm{Pr}_{e}(\zeta_{i},\zeta_{i}+\tau_{i})=P_{i}^{0}\Pr\left({X}_{i}(\bm{\xi})\geq{\zeta}_{i}+ \frac{\tau_{i}}{2}\Big|\bm{x}=\bm{\zeta}\right)\nonumber\\
&&+P_{i}^{1}\Pr\left({X}_{i}(\bm{\xi})\leq{\zeta}_{i} +\frac{\tau_{i}}{2}\Big|\bm{x}=\bm{\zeta}+\bm{\tau}_{i}\right),
\label{eq15}
\end{eqnarray}
and is bounded by the minimum error probability of the hypothesis testing problem, denoted by
$\mathcal{P}_{e}(\zeta_{i},\zeta_{i}+\tau_{i})$ which does not depend on $X_{i}(\bm{\xi})$. Thus, one obtains that
\begin{eqnarray}
&&\Pr\left(|{X}_{i}(\bm{\xi})-{x}_{i}|\geq\frac{\tau_{i}}{2}\right)\geq\nonumber\\
&&\int d\bm{\zeta}[P_{\bm{x}}(\bm{\zeta})+P_{\bm{x}}(\bm{\zeta}+\bm{\tau}_{i})]\mathcal{P}_{e}(\zeta_{i},\zeta_{i}+\tau_{i}).
\end{eqnarray}
As $\Pr\left(|{X}_{i}(\bm{\xi})-{x}_{i}|\geq\frac{\tau_{i}}{2}\right)$ is a monotonically decreasing function
of $\tau_{i}$, a tighter bound can be obtained if we fill the valleys of the right-hand side as a function
of $\tau_{i}$. Denoting this valley-filling operation as
$\mathcal{V}f(\tau_{i}) \equiv \max_{\eta\geq0}f(\tau_{i}+\eta)$,
with which one obtains the Ziv-Zakai bounds for vector parameter estimation:
\begin{eqnarray}
|\bm{\Sigma}_{\bm{\epsilon}}|&\geq&\sum_{i=1}^{d}\int_{0}^{\infty}d\tau_{i}\frac{\tau_{i}}{2}
\mathcal{V}\int d\bm{\zeta}[P(\bm{\zeta})+P(\bm{\zeta}+\bm{\tau}_{i})]\nonumber\\
&\times&\mathcal{P}_{e}(\zeta_{i},\zeta_{i}+\tau_{i}).
\end{eqnarray}

Another version that relates the mean-square error to an equally-likely-hypothesis-testing problem:
$P_{i}^{0}=P_{i}^{1}=1/2$ follows as
\begin{eqnarray}
|\bm{\Sigma}_{\bm{\epsilon}}|&\geq&\sum_{i=1}^{d}\int_{0}^{\infty}d\tau_{i}{\tau_{i}}\mathcal{V}\int d\bm{\zeta}\min\{P_{\bm{x}}(\bm{\zeta}),P_{\bm{x}}(\bm{\zeta}+\bm{\tau}_{i})\}\nonumber\\
&\times&\mathcal{P}^{el}_{e}(\zeta_{i},\zeta_{i}+\tau_{i}),
\end{eqnarray}
where $\mathcal{P}^{el}_{e}(\zeta_{i},\zeta_{i}+\tau_{i})$ denotes the minimum error probability for
equally likely hypotheses. If $P_{\bm{x}}(\bm{\zeta})$ is a uniform window, two bounds are
equivalent. From Eq.~(\ref{eq15}), we define $d$ pairs of projectors:
$\Pi_{i}^{0}+\Pi_{i}^{1}=\mathbb{I}$ which denote that
$\Pr\left({X}_{i}(\bm{\xi})\geq{\zeta}_{i}+ \frac{\tau_{i}}{2}\Big|\bm{x}=\bm{\zeta}\right)=\tr{\rho_{\bm{\zeta}}\Pi_{i}^{1}}$
and $\Pr\left({X}_{i}(\bm{\xi})\leq{\zeta}_{i}+ \frac{\tau_{i}}{2}\Big|\bm{x}=\bm{\zeta}+\bm{\tau}_{i}\right)=\tr{\rho_{\bm{\zeta}+\bm{\tau}_{i}}\Pi_{i}^{0}}$.
Then, Eq.~(\ref{eq15}) can be rewritten as
$\textrm{Pr}_{e}(\zeta_{i},\zeta_{i}+\tau_{i})=P_{i}^{0}+\tr{\Gamma_{i}\Pi_{i}^{0}}$ with
$\Gamma_{i}=P_{i}^{1}\rho_{\bm{\zeta}+\bm{\tau}_{i}}-P_{i}^{0}\rho_{\bm{\zeta}}$.
We can write $\Gamma_{i}$ in terms of its eigenvalues, which may be positive and negative:
$\Gamma_{i}=\sum_{j}\gamma_{i}^{j}|j\rangle_{i}\langle j|$, and the minimum error probability is
obtained by choosing $\Pi_{i}^{0}=\sum_{j:\gamma_{i}^{j}\leq0}|j\rangle_{i}\langle j|$:
\begin{eqnarray}
\mathcal{P}_{e}(\zeta_{i},\zeta_{i}+\tau_{i})
=\frac{1}{2}-\frac{1}{2}||P_{i}^{1}\rho_{\bm{\zeta}+\bm{\tau}_{i}}-P_{i}^{0}\rho_{\bm{\zeta}}||,
\end{eqnarray}
where $||\cdots||$ denotes to the trace norm.
The minimum error probability for equally likely hypotheses is
\begin{eqnarray}
\mathcal{P}_{e}^{el}(\zeta_{i},\zeta_{i}+\tau_{i})\geq\frac{1}{2}-\frac{1}{2}\sqrt{1-F^{2}(\rho_{\bm{\zeta}},\rho_{\bm{\zeta}+\bm{\tau}_{i}})},
\end{eqnarray}
where $F(\rho,\sigma)=\tr{\sqrt{\rho^{1/2}\sigma\rho^{1/2}}}$ refers to the Brue's fidelity between two
density matrices $\rho$ and $\sigma$, and the equal holds for pure states.
Finally, we obtain a lower bound for vector parameter estimation:
\begin{eqnarray}
|\bm{\Sigma}_{\bm{\epsilon}}|&\geq&\sum_{i=1}^{d}\int_{0}^{\infty}d\tau_{i}\frac{\tau_{i}}{2}\mathcal{V}\int d\bm{\zeta}\min\{P_{\bm{x}}(\bm{\zeta}),P_{\bm{x}}(\bm{\zeta}+\bm{\tau}_{i})\}\nonumber\\
&\times&\left(1-\sqrt{1-F^{2}(\rho_{\bm{\zeta}},\rho_{\bm{\zeta}+\bm{\tau}_{i}})}\right),
\label{a1}
\end{eqnarray}
which we call a quantum Ziv-Zakai bound for vector parameter estimation.
Another slightly tighter version
\cite{zzeb} of QZZB for multiple parameters case is written as
\begin{eqnarray}
|\bm{\Sigma}_{\bm{\epsilon}}|&\geq&\sum_{i=1}^{d}\mathcal{V}\int_{0}^{\infty}d\tau_{i}\frac{\tau_{i}}{2}\int d\bm{\zeta}[P_{\bm{x}}(\bm{\zeta})+P_{\bm{x}}(\bm{\zeta}+\bm{\tau}_{i})]\nonumber\\
&\times&\left(1-\sqrt{1-4P_{i}^{0}P_{i}^{1}F^{2}(\rho_{\bm{\zeta}},\rho_{\bm{\zeta}+\bm{\tau}_{i}})}\right).
\label{bound2}
\end{eqnarray}
These two bounds are equivalent when the vector parameter has a uniform prior
distribution as presented in Eq.~(\ref{uniform}).

\end{document}